\documentclass[preprint]{aastex}

\usepackage{verbatim}
\usepackage{natbib}

\newcommand{\derl}[2]{d #1/d #2}
\newcommand{\pder}[2]{\left(\frac{\partial #1}{\partial #2}\right)}
\newcommand{\pdernp}[2]{\frac{\partial #1}{\partial #2}}
\newcommand{\pderl}[2]{\partial #1/\partial #2}

\begin{document}
\title{Circumstellar Shell Formation in Symbiotic Recurrent Novae}
\author{Kevin Moore\altaffilmark{1} and Lars Bildsten\altaffilmark{1,2}}

\altaffiltext{1}{Department of Physics, Broida Hall, University of California, Santa Barbara, CA 93106, USA}
\altaffiltext{2}{Kavli Institute for Theoretical Physics, Kohn Hall, University of California, Santa Barbara, CA 93106, USA}

\begin{abstract}
We present models of spherically symmetric recurrent nova shells interacting with circumstellar material in a symbiotic system composed of a red giant expelling a wind, and a white dwarf accreting from this material. Recurrent nova eruptions periodically eject material at high velocities ($\gtrsim 10^3$ km/s) into the red giant wind profile, creating a decelerating shock wave as circumstellar material is swept up. High circumstellar material densities cause the shocked wind and ejecta to have very short cooling times of days to weeks. Thus, the late time evolution of the shell is determined by momentum conservation instead of energy conservation. We compute and show evolutionary tracks of shell deceleration, as well as post-shock structure. After sweeping up all the red giant wind, the shell coasts at a velocity $\sim 100$ km/s, depending on system parameters. These velocities are similar to those measured in blue-shifted circumstellar material from the symbiotic nova RS Oph, as well as a few Type Ia supernovae that show evidence of circumstellar material, such as 2006X, 2007le, and PTF 11kx. Supernovae occurring in such systems may not show circumstellar material interaction until the inner nova shell gets hit by the supernova ejecta, days to months after the explosion.
\end{abstract}


\section{Introduction}
%My old first paragraph:
%Type Ia supernovae (SNe Ia) are well-known cosmological distance indicators \citep{Howell11}, however their progenitors are still uncertain. The basic picture of a C/O white dwarf (WD) exploding once it accretes enough material to reach the Chandrasekhar mass ($M_{\rm ch}$) is complicated by uncertainty in the nature of the donor star \citep{Livio11}. The single degenerate (SD) scenario has a main sequence or slightly evolved donor, while the double degenerate (DD) scenario has another C/O WD as the donor - likely through a merger \citep{Iben84,Webbink84}. These two channels need not be mutually exclusive, but a significant problem for the SD channel is the lack of observed interaction between the supernova ejecta and the hydrogen expected to be in such a system.

%Lars' cleaned up first paragraph:
The range of Type Ia supernovae properties and new observations of these explosions at very early times \citep{Nugent11, Foley12} points to a diversity of progenitor types. The long-standing view (see review by \citet{Hillebrandt00}) has been that Type Ia result from a C/O white dwarf (WD) that has accreted enough material to compress the center to densities and temperatures so large that carbon fusion is ignited as an uncontrollable runaway that leads to an explosion. %and uncontrollably runs away to explosion. 
This requires the WD to nearly reach the Chandrasekhar mass ($M_{\rm ch}$), strongly constraining the binary evolution scenarios \citep{Livio11}. The single degenerate (SD) scenario has a main sequence or slightly
evolved donor, while the double degenerate (DD) scenario has another C/O WD as the donor - likely through a merger \citep{Iben84,Webbink84}. These two channels may both occur, but a significant problem for the SD channel as a dominant mechanism is the lack of observed interaction between the supernova ejecta and the hydrogen expected to be present in such a system.

Recent observations have shed light on possible progenitors via evidence of interaction between the supernova ejecta and circumstellar material (CSM). Samples of SNe Ia show a bias towards blue shifted spectral features, such as Na absorption, which are likely outflows from the progenitor systems themselves \citep{Sternberg11}. Previously reported Type IaÕs showing evidence of CSM include SNe 2002ic \citep{Hamuy03}, 2005gj \citep{Aldering06}, 2006X \citep{Patat07}, 2007le \citep{Simon09}, and PTF 11kx (Dilday et al., 2012). We note that there are alternative interpretations for 02ic as a Type Ic \citep{Benetti06} and 05gj as having an LBV progenitor \citep{Trundle08}. Events such as SN 2006X and PTF 11kx show multiple blue shifted absorption features which may be interpreted as previously ejected material from successive recurrent nova outbursts in a symbiotic system such as RS Oph \citep{Patat07,Patat11b}.

%\subsection{Background on novae}
A nova is a thermonuclear runaway (TNR) of accreted hydrogen on a WD that ejects material from the WD's hydrogen burning shell \citep{Bode10}. Simulations indicate that the TNR happens on a timescale ranging from days to years \citep{Yaron05}, varying with the three main parameters of an accreting WD: the mass and temperature of the WD, along with the accretion rate onto the WD \citep{Townsley04}. Theoretical models also indicate that novae recur, with a recurrence time that is sensitive to the aforementioned parameters. Novae with more than one recorded outburst are said to be recurrent novae. 

%We take the values of RS Oph as our fiducial values in the analysis that follows. Specifically, a recurrence time of $20\ {\rm yr}$, an ejected mass of $10^{-7}\ M_\odot$, an ejecta velocity of $5000$ km/s, and a wind mass-loss rate from the RG donor of $10^{-7}\ M_\odot\ {\rm yr^{-1}}$ \citep{Sokoloski06,Rupen08}.

%\subsection{Novae in symbiotic systems}
Recurrent novae can occur in both short period binaries and long period binaries \citep{Anupama08}. Short period binaries, such as T Pyx, can be cataclysmic variable (CV) systems ($P_{\rm orb}\sim 1$ hr) where a WD accretes matter from a star that overflows its Roche lobe. In long-period systems, such as RS Oph, the WD is in a wider binary ($P_{\rm orb}\sim 1$ yr) that is accreting from a wind ejected from the evolved donor, eg. a red giant. This type of nova is also referred to as a symbiotic recurrent nova (SyRNe). There are also symbiotic novae (SyNe) that have not been observed to recur (eg. AG Peg and RR Tel). Novae are uncommon among symbiotic systems: in $\sim 200$ symbiotics known there are $9$ SyNe and $4$ SyRNe \citep{Mikolajewska10}. The accretion efficiencies are quite different between recurrent novae in short-period and long-period systems. In a short-period recurrent nova, there is little circumstellar material (CSM). In a symbiotic recurrent nova, the wind accretion efficiency is so low ($\sim 1-10\%$) that most of the stellar wind is not accreted by the WD and remains as CSM. 

%\subsection{Evolution of symbiotic binaries (feasibility of accreting to $M_{\rm ch}$)}
SyRNe, especially RS Oph, have previously been investigated as possible SNe Ia progenitors \citep{Hachisu01,Wood-Vasey06,Hernanz08,Justham08,Walder10,Patat11}. Detections of circumstellar material in RS Oph \citep{OBrien06, Sokoloski06}, along with observations of spectral features blue-shifted by $30-50$ km/s during outbursts \citep{Iijima09, Patat11b} point to the existence of shells of material around the system with expansion velocities between the wind velocity, $\approx 10-20$ km/s, and the ejecta velocity, $\approx 3500$ km/s \citep{Buil06, OBrien06}.

%Move to conclusions perhaps?
%If a SyRN system is to produce a Type Ia SN, then the initial C/O WD needs to accrete enough mass to reach $M_{\rm ch}$. C/O WDs have a maximum initial mass of $\sim 1.1\ M_\odot$, so it would need to accrete roughly $0.3\ M_\odot$ from the donor's wind. Assuming an accretion/explosion efficiency of $10\%$, then the donor star would need to lose at least $3\ M_\odot$ in winds. These high amounts of mass loss can occur for very massive stars (ref?), but would make these events quite rare from the stellar mass distribution alone.

In this paper, we examine the consequences of recurrent novae on the circumbinary environment. In \S \ref{sec:kinematics} we show that the dense CSM in a symbiotic system leads to short radiative cooling times of the ejecta and shocked wind, making the kinematics of the nova shell determined by conservation of momentum. In \S \ref{sec:const_p}, we derive the equations of motion for the spherically symmetric evolution of the nova shell and provide example realizations. The time-dependent post-shock structure of the decelerating ejecta and swept up wind is discussed in \S \ref{sec:pss}, along with possible instabilities in the ejecta and their effect on the long term evolution of nova shells. Finally, \S \ref{sec:sne} discusses the implications of our model on potential supernovae in symbiotic systems.

%For CVs such as classical novae with long recurrence times there is evidence of heavy element dredge up in the spectra of nova outbursts, making it unlikely that the WD is accumulating mass. This dredge up is thought to be due to accreted H diffusing into the interior of the WD so that the runaway burning during the nova carries off some of the underlying He/C/O. Thus these systems are not good candidates for Type Ia SNe. In the case of symbiotic recurrent novae, if the recurrence times are short enough that the H doesn't have much time to diffuse into the inner C/O, then the WD could in principle grow in mass. 

%Another possible explosion mechanism in such systems is through a He shell flash in the accumulated He ashes underneath the outer accreted H. If this shell flash were to go dynamical then there is the possibility for shock focusing in the interior to detonate a sub-$M_{\rm ch}$ WD (MPA group) in a Type Ia-like explosion.

\section{Kinematics of the nova shell \label{sec:kinematics}}
An important difference between the evolution of nova ejecta in symbiotic systems versus those in cataclysmic variable systems is the large amount of CSM that the ejecta will interact with in a symbiotic system. 
The dense CSM around a symbiotic decelerates the nova ejecta and decreases the radiative cooling time by many orders of magnitude from the $>10^4$ years typical of nova shells interacting with the interstellar medium \citep{Moore11}. These short cooling times will cause momentum (rather than energy) conservation to determine the kinematics of the nova shell after a few weeks.

%There are several important timescales influencing the evolution of the nova remnant. The timescale of the nova mass ejection itself, $t_{\rm ej}\approx 5$ days for an RS Oph-like system \citep{Bode06}. The timescale to sweep up an amount of wind equal to the ejecta mass ($t_{\rm sweep}$), the timescale for the shocked ejecta to cool ($t_{\rm cool,ej}$), the timescale for the shocked wind to cool after it has been swept up ($t_{\rm cool,w}$), and the timescale for the remnant to decelerate ($t_{\rm decel}$).
%In contrast to supernova remnants where the energy-conserving self-similar Sedov phase is of great importance, here the evolution is largely determined by the momentum conserving phase.

Although there is evidence that SyRNe such as RS Oph exhibit asymmetric mass ejection \citep{OBrien06,Rupen08}, we consider a spherically symmetric model for simplicity. %(partial solid angle effects can be easily added at the end). 
Between recurrent novae (with period $t_{\rm rec}$) the red giant (RG) donor is ejecting mass at a rate $\dot{M}_w$ and velocity $v_w$, creating a density profile
\begin{equation}
\rho_w(r) = \frac{\dot{M}_w}{4\pi v_w r^2} = 2.2\times 10^{-14}\ {\rm g\ cm^{-3}} \left(\frac{\dot{M}_w}{10^{-6}\ M_\odot /{\rm yr}}\right) \left(\frac{v_w}{10\ {\rm km/s}}\right)^{-1} \left(\frac{r}{1\ {\rm AU}}\right)^{-2},
\end{equation}
out to a distance $r_{\rm max} = 6.3\times 10^{14}\ {\rm cm}\left(v_w/ 10\ {\rm km\ s^{-1}}\right) \left(t_{\rm rec}/20\ {\rm yr}\right)$ immediately before the next nova. This density profile will be perturbed by the presence of the accreting WD as shown in the simulations by \citet{Walder10}, but remains nearly axisymmetric for slower wind velocities ($v_w < 20$ km/s). Each nova ejects a mass $M_{\rm ej}$, which we scale as $M_{\rm ej} = f \dot{M}_w t_{\rm rec}$ where $f$ is a measure of both the accretion and explosion efficiency. We take $f=0.1$ as our fiducial, as various simulations show effective accretion rates between 10\% and 2\% of $\dot{M}_w$ \citep{Walder08,Walder10} with $\sim 90\%$ of the accreted material being ejected during a nova for RS Oph-like systems \citep{Hachisu01}. 

%Here is where discussion of timescales needs to go. Have to motivate momentum conserving phase in the long-term. A few ways to do this - could go back to spherical symmetry case and ask what the cooling time is at a certain time in the evolution (eg. ~10 days). We could keep doing what we're doing here and cut out the core of the density profile so that interaction starts in the first 1-2 days (~$R_s \sim 0.5$ AU). 

%The early-time evolution of the nova ejecta (on the timescale of the ejection event itself, $t_{\rm ej}\sim 7$ days) is a complex hydrodynamic process \citep{Vaytet08, Vaytet11} which we do not attempt to model. Since there's an energy and momentum input during the first few days of evolution, we start the free evolution at a time 

Our model for the evolution of the nova ejecta is motivated by observations and simulations of outbursts in RS Oph. Novae with short recurrence times also have short mass-ejection timescales \citep{Yaron05}, $\sim 1-5$ days, for recurrent novae such as RS Oph. Models of novae by \citet{Shen09} also show that short recurrence times require finely tuned mass-accretion rates. The mass-loss rate of the RG in RS Oph is $\sim 10^{-6}\ M_\odot/{\rm yr}$ \citep{Rupen08}, which is consistent with a $20$-year recurrence time given an accretion efficiency of $f=0.1$. There are a range of measurements of the ejecta mass itself. The ejecta mass of the 1985 outburst was measured to be $\sim 10^{-6}\ M_\odot$ \citep{Obrien92}. \citet{Sokoloski06} infer the ejecta mass of the 2006 outburst to be $\sim 10^{-7}\ M_\odot$ due to the quick onset of a Sedov-Taylor phase, while \citet{Vaytet11} estimate it as $(2-5)\times 10^{-7}\ M_\odot$ from long-term simulations of the x-ray emission. Theoretical light curves of the 2006 nova by \citet{Hachisu07} are best fit by an ejecta mass of $(2-3)\times 10^{-6}\ M_\odot$. The inferred ejecta mass in our model, neglecting the $\sim 10\%$ of accreted material may remain on the WD \citep{Hachisu01}, is $M_{\rm ej} = f \dot{M}_w t_{\rm rec} \approx 2\times 10^{-6}\ M_\odot$, on the high side of estimates for RS Oph. Most measurements of the ejecta velocity are $v_{\rm ej} = 3000-3500$ km/s \citep{Hjellming86,Buil06}, but models from \citet{Vaytet11} argue it is much higher, $6000-10000$ km/s.

Early-time evolution of the ejection event is complex, requiring hydrodynamic wind-wind interactions \citep{Vaytet07, Vaytet11}, which we do not attempt to model. The 2006 outburst of RS Oph showed rapidly decelerating ejecta matching the self-similar Sedov-Taylor phase $3-10$ days after the beginning of the outburst \citep{Sokoloski06,Bode06}, and quickly transitioning to the momentum-conserving phase after $\sim 14$ days \citep{Bode06,Rupen08}. These observations indicate that the ejecta sweeps up enough mass in $\sim 3$ days to get reverse-shocked and be in the self-similar phase. The cooling time of the postshock material is thus $\sim 14$ days in order to make the transition to a momentum conserving phase.

The simplest model is to have the nova ejecta (here taken to be ejected all at once) coast into the $\rho_w$ profile described above until it has swept up mass equal to itself, at at time $t_{\rm sweep} = 0.3$ days given our fiducials of $\dot{M}_w = 10^{-6}\ M_\odot$/yr, $v_w = 10$ km/s, $v_{\rm ej} = 3000$ km/s, and $t_{\rm rec} = 20$ yrs. A slightly more realistic model is to remove the core of the wind profile so the ejecta starts encountering mass after it has travelled roughly the orbital separation, $a \sim 0.5$ AU. Doing so increases $t_{\rm sweep}$ to $3.0$ days. After this time the shell is in the self-similar Sedov-Taylor phase until radiative cooling makes the energy-conserving assumption invalid.

While in the Sedov-Taylor phase, the position and velocity of the forward shock are given by \citep{Chevalier82a}
\begin{eqnarray}
R_s &=& 9.8\ {\rm AU} \left(\frac{f}{0.1}\right)^{1/3} \left(\frac{t_{\rm rec}}{20\ {\rm yrs}}\right)^{1/3} \left(\frac{v_w}{10\ {\rm km/s}}\right)^{1/3} \left(\frac{v_{\rm ej}}{3000\ {\rm km/s}}\right)^{2/3} \left(\frac{t}{5\ {\rm days}}\right)^{2/3} \\
v_s &=& 2300\ {\rm km/s} \left(\frac{f}{0.1}\right)^{1/3} \left(\frac{t_{\rm rec}}{20\ {\rm yrs}}\right)^{1/3} \left(\frac{v_w}{10\ {\rm km/s}}\right)^{1/3} \left(\frac{v_{\rm ej}}{3000\ {\rm km/s}}\right)^{2/3} \left(\frac{t}{5\ {\rm days}}\right)^{-1/3}.
\end{eqnarray}
The post-shock material is assumed to be fully reverse-shocked in this phase, so we do not follow any transient reverse shocks. The immediate post-shock particle density at time $t$ is given by the strong shock jump conditions \citep{Draine11}
\begin{eqnarray}
n_e = n_H = 9.1\times 10^8\ {\rm cm}^{-3} \left(\frac{\mu}{0.6}\right)^{-1} \left(\frac{f}{0.1}\right)^{-2/3} \left(\frac{\dot{M}_w}{10^{-6}\ M_\odot/{\rm yr}}\right) \left(\frac{v_w}{10\ {\rm km/s}}\right)^{-5/3} \nonumber \\ \left(\frac{v_{\rm ej}}{3000\ {\rm km/s}}\right)^{-4/3} \left(\frac{t_{\rm rec}}{20\ {\rm yrs}}\right)^{-2/3} \left(\frac{t}{5\ {\rm days}}\right)^{-4/3},
\end{eqnarray}
where $\mu$ is the mean molecular weight of the material. The post-shock temperature is
\begin{equation}
T_s = 1.4\times 10^7\ {\rm K} \left(\frac{\mu}{0.6}\right) \left(\frac{v_s}{1000\ {\rm km/s}}\right)^2.
\end{equation}
%The average particle density gives a lower limit on the particle density in the cooling post shock region,
%\begin{equation}
%n_e = n_H = 3.0\times 10^8\  {\rm cm}^{-3} \left(\frac{\dot{M}_w}{10^{-6} M_\odot/{\rm yr}}\right) \left(\frac{\mu}{0.6}\right)^{-1} \left(\frac{v_w}{10\ {\rm km/s}}\right)^{-1} \left(\frac{v_{\rm ej}}{3000\ {\rm km/s}}\right)^{-2} \left(\frac{t}{5\ {\rm days}}\right)^{-2}.
%\end{equation}
The post-shock cooling is roughly isobaric \citep{Bertschinger86} so the cooling time of the post-shock material is
\begin{equation}
t_{\rm cool} = \frac{nkT_s}{(\gamma-1)\Lambda},
\end{equation}
where $\gamma=5/3$ is the adiabatic index of the gas and $\Lambda$ is the cooling function. Using the approximation for the cooling function of $\Lambda/(n_e n_H) = 1.1\times 10^{-22}\ T_6^{-0.7}$ erg cm$^3$ s$^{-1}$ ($T_6$ is the temperature in units of $10^6$ K), valid for $10^5\ {\rm K} < T < 10^{7.3}\ {\rm K}$ \citep{Draine11}, we calculate
\begin{eqnarray}
t_{\rm cool} = 36\ {\rm days} \left(\frac{\mu}{0.6}\right)^{2.7} \left(\frac{f}{0.1}\right)^{1.8} \left(\frac{v_w}{10\ {\rm km/s}}\right)^{2.8} \left(\frac{v_{\rm ej}}{3000\ {\rm km/s}}\right)^{3.6} \nonumber \\
\left(\frac{\dot{M}_w}{10^{-6} M_\odot/{\rm yr}}\right)^{-1} \left(\frac{t_{\rm rec}}{20\ {\rm yrs}}\right)^{1.8} \left(\frac{t}{5\ {\rm days}}\right)^{0.2}.
%t_{\rm cool} = 18\ {\rm days} \left(\frac{\mu}{0.6}\right)^{0.3} \left(\frac{f}{0.1}\right)^{0.2} \left(\frac{v_w}{10\ {\rm km/s}}\right)^{1.2} \left(\frac{v_{\rm ej}}{3000\ {\rm km/s}}\right)^{2.4} \left(\frac{\dot{M}_w}{10^{-6} M_\odot/{\rm yr}}\right)^{-1} \left(\frac{t}{5\ {\rm days}}\right)^{1.8},
\end{eqnarray}
Using numerically computed cooling functions \citep{Gnat07}, rather than the power-law fit used above, reduces the cooling time to $t_{\rm cool} \approx t$ for the first $16$ days of evolution, after which $t_{\rm cool} < t$, roughly agreeing with the cooling time inferred from shock deceleration measurements of RS Oph outlined above.

\section{Momentum-conserving evolution \label{sec:const_p}}
We now derive the equation of motion for the momentum-conserving phase. As will be shown in \S \ref{sec:pss}, the rapid cooling of the shocked material at late times causes most of the ejecta to be moving at the same velocity as the shock front, $v_s$. The initial momentum of the system is split between the momentum in the ejecta, $p_{\rm ej} = f \dot{M}_w t_{\rm rec} v_{\rm ej}$ 
%(assuming all the ejecta is moving at the same velocity - show in prev section) 
and that in the wind, $p_{w} = (1-f) \dot{M}_{w} t_{\rm rec} v_{w}$. We immediately derive the final coasting velocity of the shell after it has swept up all the wind, using $p_{\rm final} = \dot{M}_w t_{\rm rec} v_{\rm coast}$, and thus 
\begin{equation}
v_{\rm coast} = f v_{\rm ej} + (1-f) v_w.
\end{equation}
This implies coasting velocities of $\simeq 100$ km/s, intermediate to both the wind and nova velocities. At a time $t$ after the nova event the ejecta is sweeping up the wind, and the total wind mass swept up when the shell is at radius $R_s(t)$,
\begin{equation}
M_{\rm sweep}(t) = \int_{v_{\rm wind} t}^{R_s(t)} 4\pi r^2 \rho_w(r) dr = \frac{\dot{M}_{w}}{v_{w}}\left(
R_s(t) - v_{w} t \right).
\end{equation}
The integration must start at $v_{\rm wind} t$ because that is the outer radius of the wind that was ejected since the nova outburst. From this, we define the column (number) density of the shell as
\begin{equation}
N = \frac{M_{\rm sweep}(t)}{4\pi \mu m_p R_s(t)^2}.
\end{equation}
%\citet{Shore96} measured the column density of the red giant wind to be $\sim 10^{21}$ cm$^{-2}$ via iron-group absorption lines in the pre-UV maximum phase, similar to the our fiducial of $8\times 10^{20}$ cm$^{-2}$.

\begin{figure}
	\plotone{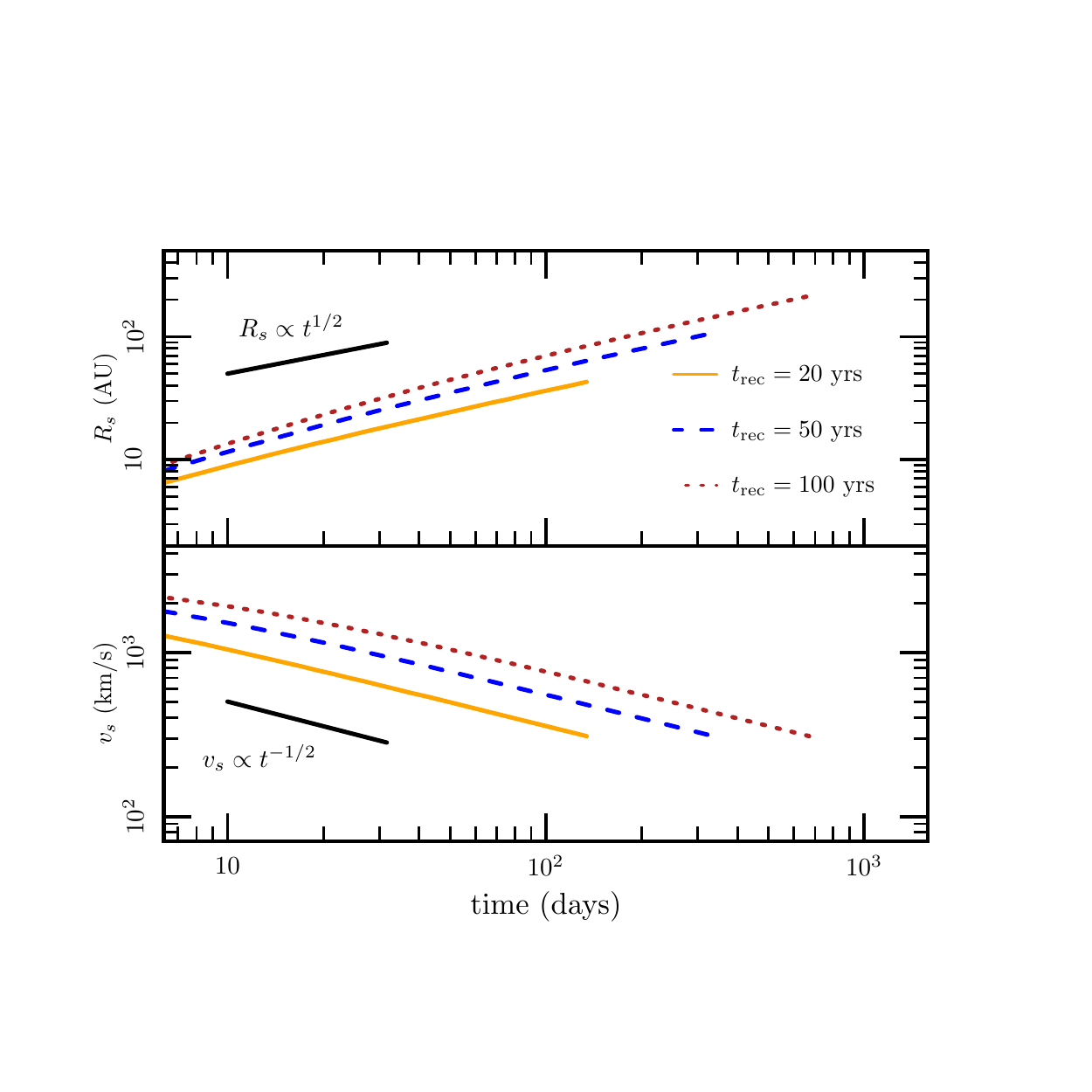}
	\caption{\label{fig:rtvt} Shock radius ($R_s$, top panel) and velocity ($v_s$, bottom panel) as a function of time in the momentum conserving phase for different recurrence times. The lines end at $t_{\rm sweep}$, once the ejecta has swept up all the wind, reaching a velocity $v_{\rm coast}$. The accretion/explosion efficiency is $f = 10^{-1}$, the RG mass loss rate is $\dot{M}_w = 10^{-6}\ M_\odot/$yr, the wind velocity is $v_w = 10$ km/s, and the ejecta velocity is $v_{\rm ej} = 3000$ km/s. This system has $v_{\rm coast} = 309$ km/s.}
\end{figure}

The equation of motion for the shell arises from conservation of momentum:
\begin{equation}
(M_{\rm sweep}(t) + f \dot{M}_{w} t_{\rm rec})v_s(t) + \left[(1-f) \dot{M}_{w} t_{\rm rec} - M_{\rm sweep}(t)\right]v_{w} = \dot{M}_{w} t_{\rm rec}\left[f v_{\rm ej} + (1-f)v_{w}\right],
\end{equation}
and thus
\begin{equation}
\ddot{R}_s = \frac{-(1-f)(v_s(t)-v_w)^2}{(1-f)(R_s(t)-v_w t) + f v_w t_{\rm rec}}.
%Still correct 1st order equation, but we used 2nd order one above in calculations (needed accelerations)
%v_s(t) = \dot{R}_s(t) = \frac{v_w (f t_{\rm rec} v_{\rm ej} + R_s(t) - v_w t)}{R_s(t) + v_w(f t_{\rm rec} - t)}.
\end{equation}
We give examples of $R_s(t)$ and $v_s(t)$ in Fig. \ref{fig:rtvt}. 
%The same trends in velocity and radius evolution appear in radio observations of RS Oph \citep{Rupen08}.

We can also obtain a simpler equation of motion in the limit of negligible wind velocity $(R_s \gg v_w t)$,
\begin{equation}
	%R(t) = -f v_w t_{\rm rec} + \sqrt{(f v_w t_{\rm rec})^2 + 2 f v_w v_{\rm ej} t_{\rm rec} t}.
	R_s(t) = f v_w t_{\rm rec} \left(\sqrt{1+\frac{2 v_{\rm ej} t}{f v_w t_{\rm rec}}} - 1\right),
\end{equation}
defining an evolution timescale
\begin{equation}
t_{\rm evol} =  \frac{f v_w t_{\rm rec}}{2 v_{\rm ej}} = 4\ {\rm days}\  \left(\frac{f}{0.1}\right) \left(\frac{v_w}{10\ {\rm km/s}}\right) \left(\frac{t_{\rm rec}}{20\ {\rm yr}}\right) \left(\frac{v_{\rm ej}}{1000\ {\rm km/s}}\right).
\end{equation}
Thus, for early times ($t \ll t_{\rm evol}$) $R_s(t) \propto t$, while at late times $R_s(t) \propto t^{1/2}$. Coincidentally, $t_{\rm evol}$ is shorter than the mass-ejection timescale of the nova, $t_{\rm ej}$, as well as $t_{\rm cool}$ in the Sedov-Taylor phase, and the momentum-conserving solution is not valid at these early times. Most of the observed shell evolution should therefore be in the $R_s(t) \propto t^{1/2}$ phase. High resolution radio observations of the 2006 outburst of RS Oph tracked the deceleration of the shell, with \citet{Rupen08} finding $v_s \propto t^{-0.52}$ in the period $14-27$ days after maximum - agreeing with the kinematics predicted by a momentum conserving phase after a brief Sedov-Taylor phase.

\begin{figure}
	\plotone{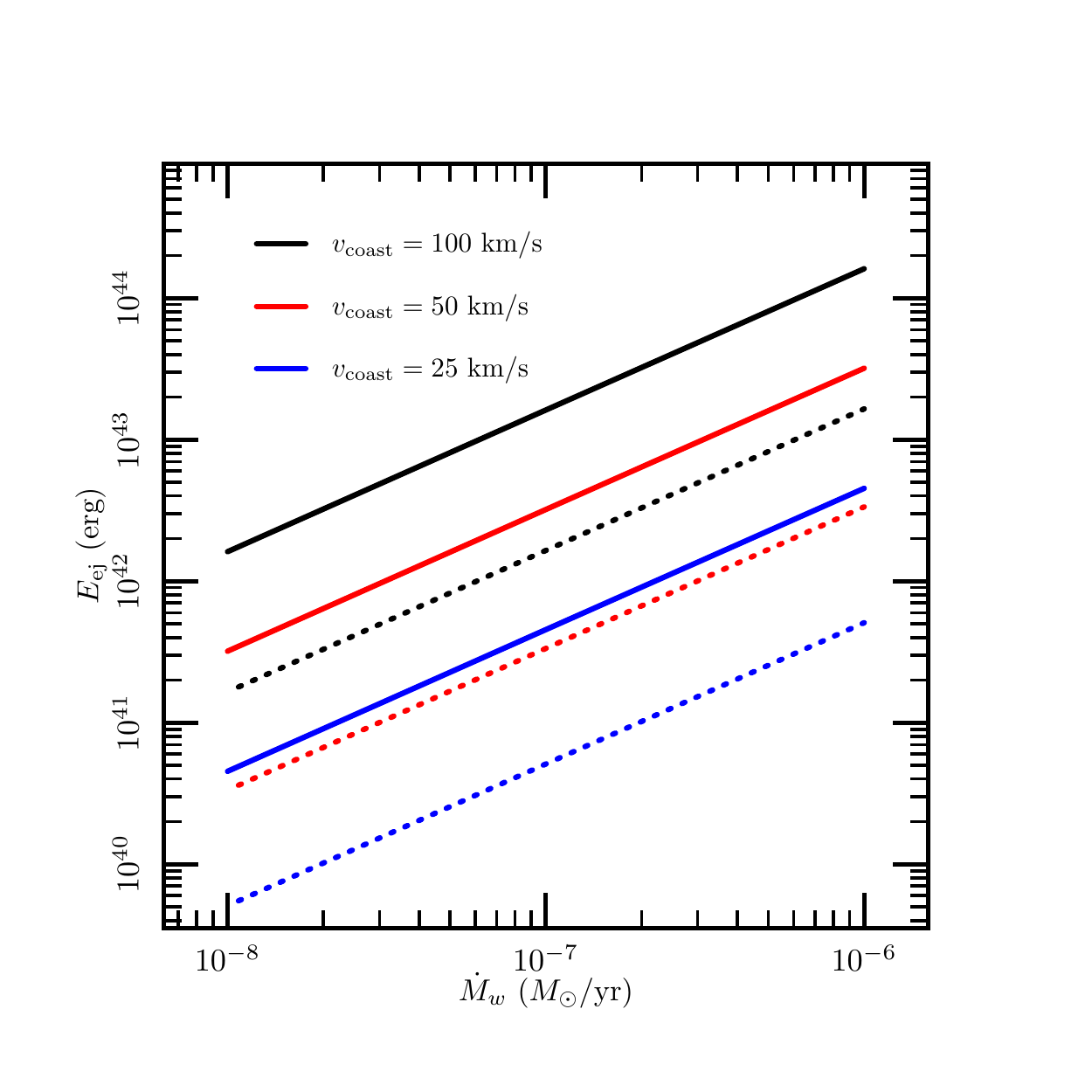}
	\caption{\label{fig:vcoast} Lines of constant $v_{\rm coast}$ calculated on the $\dot{M}_w-E_{\rm ej}$ plane. The ejecta energy is assumed to be kinetic energy dominated: $E_{\rm ej} = f\dot{M}_w t_{\rm rec} v_{\rm ej}^2/2$. Solid lines are for accretion/explosion efficiencies of $f=10^{-2}$ and dotted lines are for $f=10^{-1}$. The other model parameters have been fixed at the fiducial values of $v_w = 10$ km/s and $t_{\rm rec} = 20$ yrs.}
\end{figure}

These calculations also yield the time required to completely sweep up the wind ejected since the previous nova, $t_{\rm coast}$. From that time onward, the shell of material simply coasts outward at velocity $v_{\rm coast}$. We show in the following section that the swept up material is in a geometrically thin shell at $R_s$ and has a nearly uniform velocity throughout. In order to illustrate the resulting diversity in expected column densities, $N_{\rm col}$, and coasting velocities, we rewrite our results in terms of ejecta energy, $E_{\rm ej}$ and $\dot{M}_w$ since these are more accessible observationally. From $E_{\rm ej} = f\dot{M}_w t_{\rm rec} v_{\rm ej}^2/2$ we get
\begin{equation}
E_{\rm ej} = \frac{\dot{M}_w t_{\rm rec}}{2f}\left[v_{\rm coast} - (1-f)v_w\right]^2.
\end{equation}
From this, lines of constant $v_{\rm coast}$ are shown on the $E_{\rm ej} - \dot{M}_w$ plane in Figure \ref{fig:vcoast}. Typical ejecta energies of SyRNe are $\sim 10^{42}-10^{44}$ erg. The 1985 outburst of RS Oph had measured ejecta energies of $8\times10^{42}$ erg \citep{Bode85} and $1.1\times 10^{43}$ erg \citep{Obrien92}, while the 2010 outburst of V407 Cyg was estimated at $2\times 10^{44}$ erg \citep{Orlando12}. 

\begin{figure}
	\plotone{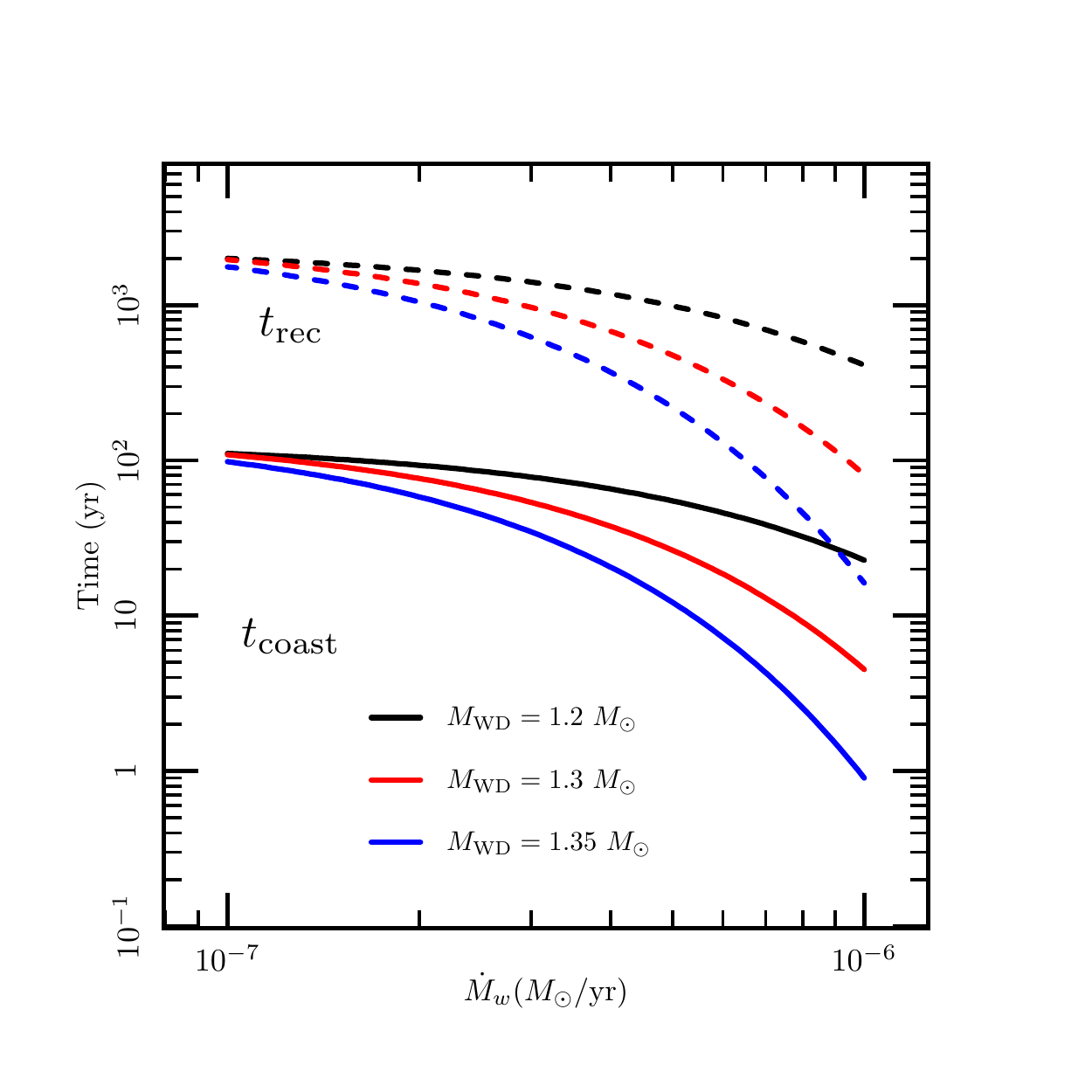}
	\caption{\label{fig:tcoast} Variation of the recurrence time, $t_{\rm rec}$, and the time to sweep up the entire wind, $t_{\rm coast}$, with $\dot{M}_w$ for novae on white dwarfs with different masses. The nova recurrence times vary with accretion rate and white dwarf mass as in \citet{Shen09}. We calculate these curves with an accretion/explosion efficiency of $f=0.1$, a wind velocity $v_{\rm wind} = 10$ km/s, and ejecta velocity $v_{\rm ej} = 5000$ km/s (thus a coasting velocity of $v_{\rm coast} = 509$ km/s). Since $t_{\rm coast}$ is always less than $t_{\rm rec}$, the shell has reached its coasting velocity (i.e. the shock has broken out of the wind nebula) long before the next nova occurs.}
\end{figure}

\begin{figure}
	\plotone{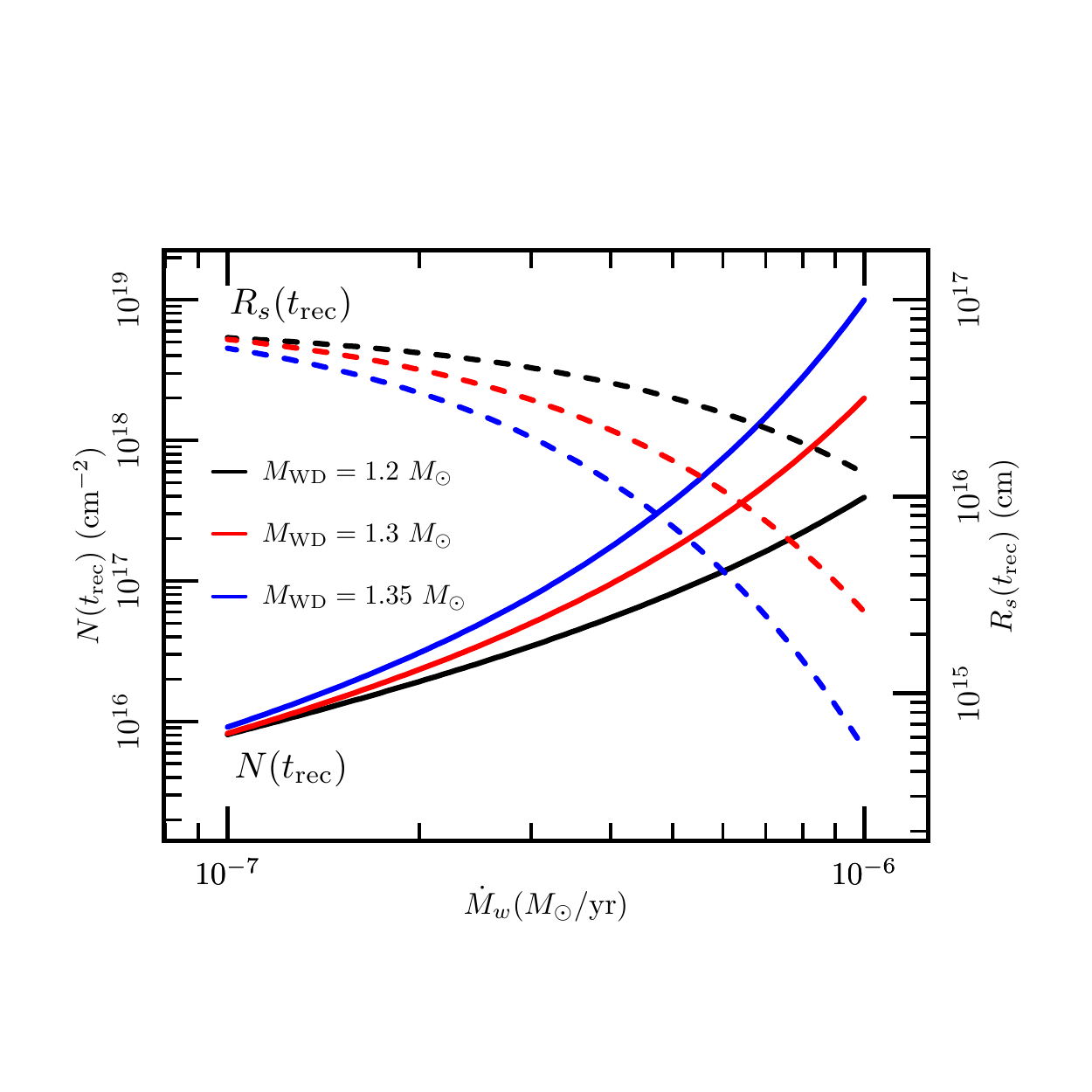}
	\caption{\label{fig:ncol} Lines of $N(t_{\rm rec})$ and $R_s(t_{\rm rec})$ for different WD masses as a function of $\dot{M}_w$. Solid lines are $N(t_{\rm rec})$ values (left axis), while dashed lines are $R_s(t_{\rm rec})$ values (right axis). The recurrence times vary with accretion rate and white dwarf mass as in \citet{Shen09}. We calculate these curves with an accretion/explosion efficiency of $f=0.1$, a wind velocity $v_{\rm wind} = 10$ km/s, and ejecta velocity $v_{\rm ej} = 3000$ km/s (thus a coasting velocity of $v_{\rm coast} = 309$ km/s).}
	
	%\caption{\label{fig:ncol} Lines of constant $n_{\rm col}t^2$ on the on the $\dot{M}_w-E_{\rm ej}$ plane. Solid lines are for accretion/explosion efficiencies of $f=10^{-2}$ and dotted lines are for $f=10^{-1}$. The other model parameters have been fixed at the fiducial values of $v_w = 10$ km/s and $t_{\rm rec} = 20$ yrs. The expected late-time evolution of the column density after an nova is $n_{\rm col} = \dot{M}_w t_{\rm rec}/4\pi (v_{\rm coast} t)^2$, thus $n_{\rm col}t^2$ should be constant at late-times. There is a sharp drop off in the necessary $E_{\rm ej}$ to provide a given $n_{\rm col}t^2$ because $v_{\rm coast}$ is proportional to $\dot{M}_w$ in the equation for $n_{\rm col}t^2$, and can go to zero for a small enough values of $\dot{M}_w$.}
\end{figure}

Nova recurrence times depend on the accretion rate and WD mass \citep{Yaron05,Shen09}. Using the calculations from \citet{Shen09}, we compute properties of the ejecta at the varying recurrence time as a function of accretion rate (for our purposes, $\dot{M}_w$). We show the resulting timescales $t_{\rm rec}$ and $t_{\rm coast}$ in Figure \ref{fig:tcoast}. We are also interested in the state of the nova ejecta when the next nova occurs. We plot the column density and the extent of the shell at $t_{\rm rec}$ in Figure \ref{fig:ncol}. The value of $N(t_{\rm rec})$ for our fiducial values is $4\times 10^{19}$ cm$^{-2}$.

%The column density of the shell is $n_{\rm col}(t) = \dot{M}_w t_{\rm rec} / 4\pi R_s(t)^2$, and at late times it can be written as
%\begin{equation}
%n_{\rm col} = \frac{ \dot{M}_w t_{\rm rec}}{4\pi (v_{\rm coast} t)^2},
%\end{equation}
%which is valid when 
%\begin{equation}
%t >> \frac{r_{\rm sweep} - v_{\rm coast} t_{\rm sweep}}{v_{\rm coast}},
%\end{equation}
%around $300\ {\rm yr}$ for our fiducials.

%At these late times, $n_{\rm col} t^2$ will thus be roughly constant and equal to
%\begin{equation}
%n_{\rm col}t^2 = 4.5\times 10^{12}\ {\rm g\ s^2/cm^2} \left(\frac{\dot{M}_w}{10^{-7}\ M_\odot/{\rm yr}}\right)\left(\frac{t_{\rm rec}}{20\ {\rm yr}}\right) \left(\frac{v_{\rm coast}}{84\ {\rm km/s}}\right)^{-2}.
%\end{equation}
%We plot lines of constant $n_{\rm col}t^2$ in fig \ref{fig:ncol} using 
%\begin{equation}
%E_{\rm ej} = \left(\frac{-v_w(1-f)}{2} \sqrt{\frac{2\dot{M}_w t_{\rm rec}}{f}} + \frac{1}{2}\sqrt{\frac{\dot{M}_w^2 t_{\rm rec}^2}{2\pi f (n_{\rm col} t^2)}}\right)^2.
%\end{equation}
%We note that there is a critical $\dot{M}_w$ value below which $v_{\rm coast}$ will be negative and unphysical for a fixed $n_{\rm col} t^2$. This critical $\dot{M}_w$ is
%\begin{eqnarray}
%\dot{M}_{w,{\rm crit}} &=& \frac{4\pi v_w^2(1-f)^2(n_{\rm col}t^2)}{t_{\rm rec}} \\
%&=& 2.2\times 10^{-7}\ M_\odot/{\rm yr}\ \left(\frac{v_w}{10\ {\rm km/s}}\right)^2 \left(\frac{1-f}{0.99}\right)^2 \left(\frac{n_{\rm col} t^2}{10^{14}\ {\rm g\ s^2/cm^2}}\right) \left(\frac{t_{\rm rec}}{20\ {\rm yr}}\right)^{-1}.
%\end{eqnarray}

%\begin{figure}
%	\plotone{vt_plot_5000_10.pdf}
%	\caption{\label{fig:vt} Shock velocity as a function of time in the momentum conserving phase for different recurrence times. The accretion/explosion efficiency is $f = 10^{-2}$. The dotted line shows the velocity once they sweep up all the wind mass, $59.9$ km/s.}
%\end{figure}

\newpage

\section{Post-shock structure and density estimates during deceleration \label{sec:pss}}
We have derived the expected column densities and coasting velocities from the simplest momentum conserving considerations. The resulting values agree with those inferred in the few clear observations of circumstellar shells \citep{Dilday12,Patat11,Patat07,Hamuy03}. However, we have not calculated the thickness, $\Delta R$, of this shell, which sets the number density that is critical to photoionization and recombination calculations \citep{Simon09}. Such estimates require a consideration of the post-shock structure in the deceleration phase.

\subsection{Cooling gas}
Calculation of post-shock structure of the nova blast wave is divided into two main parts - cooling and cooled gas. Immediately behind the shock front is the region of active radiative cooling. Since we have already established that the cooling time is short ($t_{\rm cool} \ll R_s/v_s$), we calculate the fluid structure in this phase using the steady-state approximation. The hydrodynamic equations are written in Lagrangian coordinates %, where $\derl{f}{t} = \partial f/\partial t + (u\cdot \nabla) f$.
%\begin{eqnarray}
%\der{\rho}{t} &=& -\rho(\nabla\cdot u) \label{eq:hydrodef1} \\
%\der{u}{t} &=& -\frac{1}{\rho}\nabla P \label{eq:hydrodef2} \\
%\der{P}{t} + \frac{\gamma P}{\rho}\der{\rho}{t} &=& -(\gamma -1)\Lambda \label{eq:hydrodef3}.
%\end{eqnarray}
and we work in spherical coordinates so that the radial fluid velocity is represented by the single variable $u_r$. In the steady state approximation %the explicit time derivatives ($\partial f/\partial t$) in eqns. \ref{eq:hydrodef1}-\ref{eq:hydrodef3} go to zero and 
we are left with the single independent variable $r$, indicating the radius from the explosion center. This yields three first-order differential equations:
\begin{eqnarray}
\pdernp{\rho}{r} &=& \frac{-(\gamma-1)\Lambda/u_r -2\rho^2u_r^2/r}{u_r^2-\gamma P/\rho}, \\
\pdernp{u_r}{r} &=& -\frac{u_r}{\rho}\pder{\rho}{r}-\frac{2\rho u_r}{r}, \\
\pdernp{T}{r} &=& \frac{-(\pderl{P}{\rho})_T(\pderl{\rho}{r})-\rho u_r (\pderl{u_r}{r})}{(\pderl{P}{T})_\rho}.
\end{eqnarray}

The post-shock conditions at $R_s$ come from the strong shock jump equations: $\rho_1 = (\gamma+1)\rho_w(R_s)/(\gamma-1)$, $kT_1 = 2(\gamma+1)\mu m_p v_s^2/(\gamma+1)$, and the ideal gas equation of state $P_1 = \rho_1 k T_1/(\mu m_p)$. Here we use non-equilibrium cooling functions from \citet{Gnat07} since cooling can be rapid enough to throw ion abundances out of equilibrium. We integrate the fluid equations from $R_s$ going in until the cooling rate drops off at $T_c=10^4\ K$ and radiative cooling is no longer important. The cooling time is always much less than the age of the shell, so the shocked wind also follows the momentum conserving evolution derived above. As the post-shock gas cools, it slows down relative to the shock front and increases in density.

We can estimate the density increase in the cold gas by noting that the cooling is roughly isobaric. The post-shock pressure is $P_1 \approx \rho_0 v_s^2$, which must equal the pressure of the gas after it has cooled, $P_c = \rho_c k T_c / \mu m_p$ ($c$ subscript indicates values in the cold gas). In terms of the thermal velocity of the cold gas, $v_{\rm th} = \sqrt{3 k T_c / \mu m_p} = 21\ {\rm km/s}$, this gives
\begin{equation}
\frac{\rho_c}{\rho_0} \approx \left(\frac{v_s}{v_{\rm th}}\right)^2 = 20 \left(\frac{v_s}{100\ {\rm km/s}}\right)^2.
\end{equation}
Note the shock velocity is much slower than the ejecta velocity for all but the earliest evolutionary phases. A significant density increase in the cooled gas is also seen in Fig. 5 of \citet{Vaytet11}, although the simulation shown is at a much higher ejecta velocity ($10^4$ km/s) and earlier time ($3$ days) so we can only make a qualitative comparison. From our derived density contrast, we can also estimate the thickness of the cold gas via $4\pi R_s^2 \rho_c \Delta r \approx M_{\rm sweep}(t)$ and thus
\begin{equation}
\frac{\Delta R}{R_s} \approx \left(\frac{v_{\rm th}}{v_s}\right)^2.
\end{equation}
The cold gas is therefore a thin, dense shell as compared to the post-shock gas at $R_s$.

%From the 1D fluid equations and the strong-shock jump conditions, the pressure as a function of flow velocity $u$ (relative to the velocity of the shock front) is
%\begin{equation}
%	P(u) = \rho_0 u_0\left(u_0 - u \right).
%\end{equation}
%In the post-shock region the flow velocity varies from $u_0/4$ to $0$ while the shocked wind is cooling. As the pressure approaches $P_c = \rho_0 u_0^2$ the density approaches $n_c \approx \rho_0 u_0^2/k T_c$ (compared to the post shock value of $n_c \approx 4 \rho_0/\mu m_p$). This layer of cooled material will have a thickness 
%\begin{equation}
%l_{\rm cold} \sim \mathcal{M}_{\rm cold}^2 R = \left(\frac{\mu m_p \dot{R}}{k T_c}\right) R,
%\end{equation} 
%where $\mathcal{M}_{\rm cool}$ is the mach number in the cooled material. There will be a small residual velocity gradient over this length scale which will cause the layer to further expand after the ejecta sweeps up the entire wind mass.

\subsection{Cold gas evolution}
\begin{figure}
	\centering
	\includegraphics[scale=1.2]{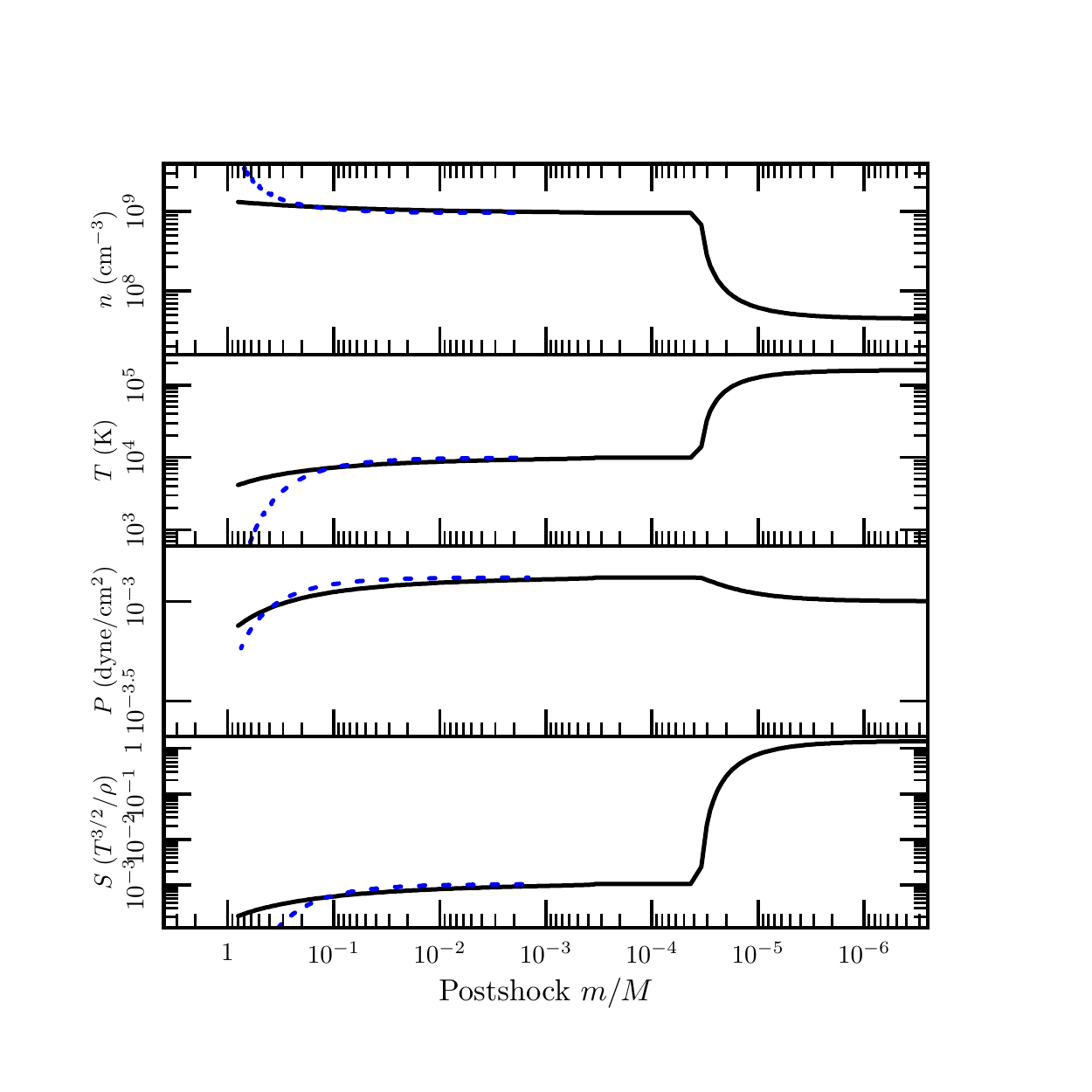}
	\caption{\label{fig:ps_struc} Number density, pressure, temperature, and entropy (via $T^{3/2}/\rho$) evolution versus mass coordinate behind the shock. The black lines are from a calculation using the separation of variable method following \citet{Bertschinger86}, while the blue dashed lines are from the thin-shell calculation described in this paper. Gas is cooling from the shock front until $m/M \approx  4\times 10^{-5}$, while gas behind that has already cooled (from a slightly different shock temperature). The gas near the rear of the shell is at an even lower temperature since it has undergone significant adiabatic expansion. These structure plots show when a nova shell has just swept up all the wind (at $t_{\rm coast} = 1.1$ yrs) in a system with $f=0.1$, $v_{\rm ej} = 1000$ km/s, $\dot{M}_w = 10^{-6}\ M_\odot/$yr, $v_w=10$ km/s, and $t_{\rm rec} = 20$ yrs.}
\end{figure}

Behind the cooling layer is the layer of cold, swept up material. The evolution of this material depends on the secular evolution of the shock front, so the steady-state approximation breaks down. One method to obtain the  structure in the swept up matter is to assume that the deceleration of the shock front is quickly transmitted to all the gas, so that the internal pressure gradient is determined by the evolution of the shock front, $g\equiv \derl{v_s}{t} = -dP/\rho dr$. By switching to mass coordinate behind the shock, $m$, we can write
\begin{equation}
\frac{dP}{dm}=\frac{g}{4\pi r^2},
\end{equation}
where $m$ is the mass outside of radius $r$, $m = \int_r^{R_s} 4\pi r^2 \rho(r) dr$, so $m/M$ is a mass coordinate measured from inside the shock (so that the shock front is at $m/M=0$) and $g$ is negative since the shell is decelerating. The entropy of each mass element of wind gets frozen after radiative cooling has ceased. We find the entropy structure, $S(m)$, by using the steady state cooling calculation described in the previous section. Finally, we assume that most of the mass is concentrated in a thin shell near the shock front ($r=R_s$), and calculate the pressure structure from hydrostatic balance,
\begin{eqnarray}
\int_{P(0)}^{P(m)} dP &=& g \int_0^m \frac{dm}{4\pi r^2}\\
\Rightarrow P(m) &=& P(0) + \frac{gM}{4\pi R_s^2}\left(\frac{m}{M}\right).
\end{eqnarray}
Combining this with the input entropy profile and the adiabatic relations $\rho \propto (P/P_0)^{3/5}$ and $T \propto (P/P_0)^{2/3}$ gives us the full post-shock structure (except for the velocity), shown in blue lines in Fig. \ref{fig:ps_struc}. We compare this calculation to that in \citet{Bertschinger86} which assumes a self-similar solution in the cold-gas. That solution does not reproduce the post-cooling entropy of the shocked wind at early times because it uses a single power-law evolution in time, while the solution is transitioning from a radiative shock to a momentum-conserving one. The thin-shell approximation is more accurate for the regime where the outward moving shock has not yet reached $R_s \propto t^{1/2}$.

\subsection{Instabilities and long-term evolution \label{sec:instabilities}}
\citet{Chevalier82b} investigated instabilities of radiative shock waves in the ISM. Their analysis showed than for a cooling function $\Lambda \propto T^\alpha$, oscillatory cooling instabilities appear for $\alpha < 0.4$ (fundamental mode) and $\alpha < 0.8$ (first overtone). Recall that a fit to the equilibrium cooling function at relevant temperatures had $\Lambda \propto T^{-0.7}$, indicating an instability. We note, as do \citet{Chevalier82b}, that non-equilibrium effects significantly alter the cooling function \citep{Gnat07} so a true time-dependent calculation is necessary to determine the presence of instabilities in such radiative shocks. Such time-dependent calculations have been carried out for 1D piston-driven radiative shocks \citep{Innes87, Gaetz88}, showing that shock velocities above $v_s \approx 150$ km/s are `overstable', having oscillations in shock position relative to the piston location. Thus, steady-state shock calculations are not suitable for investigating the luminosities and spectra during the cooling phase of of a radiative shock in our model.

\section{Supernovae in symbiotic systems \label{sec:sne}}
Supernovae occurring in such systems should show signals of interaction with CSM. Donor star mass loss rates in symbiotic systems vary between first-ascent RGB stars, $10^{-7} - 10^{-5} \ M_\odot/{\rm yr}$ \citep{Seaquist90}, and extreme asymptotic giant branch (AGB) stars. 
%My old text:
%There appears to be a spectrum of CSM amounts in interacting Ia's with events such as 2002ic estimated as having $\sim 1\ M_\odot$ of CSM \citep{Wood-Vasey06, Nomoto05} and PTF11kx having much less \citep{Dilday12}, although possibly more than could be provided by a typical RGB star wind.
%Ben's text:
The SNe Ia that have shown prominent interaction with CSM vary in strength from SN 2005gj \citep{Aldering06} to SN 2002ic \citep{Wood-Vasey06, Nomoto05, Hamuy03} to PTF 11kx \citep{Dilday12}, implying a continuum of progenitor properties such as CSM mass accumulation or time since the most recent nova event. The prominent interaction seen in these SNe Ia appear to require a mass of CSM greater than can be provided by an RGB wind on the time-scale of ~100 years (suggesting a recent AGB phase), but less prominent interaction with the wind/ nova shells of an RGB star is also detectable.

Assuming a supernova could go off at any phase within the recurrent nova cycle, there will likely be some CSM inside the inner nova cavity that the supernova will quickly sweep up. Early-time radio observations have looked for CSM interactions in Type Ia's assuming companion wind \citep{Panagia06}, and placed stringent mass loss rates ($\dot{M}_w\lesssim 3\times 10^{-8}\ M_\odot/{\rm yr}$) on winds in the progenitor systems assuming uniform progenitors across a set of Ia's. Individual Ia's used in that study had typical mass-loss constraints $\dot{M}_w\lesssim 10^{-7} - 10^{-6}\ M_\odot/{\rm yr}$. Recently, the excellent early-time observations of 2011fe indicate $\dot{M}_w\lesssim 6\times 10^{-10}\ M_\odot/{\rm yr}$ for that system \citep{Chomiuk12}.

As pointed out in \citet{Wood-Vasey06}, these early time CSM interactions could be avoided if a previous nova swept out a cavity that has been refilled for less than $t_{\rm rec}$. Assuming an ejecta velocity of $v_{\rm SN} = 10^4$ km/s for the supernova, then in the first $t$ days it could sweep up a wind mass that had been ejected for the last $t_{\rm SN}$ years, where
\begin{equation}
t_{\rm SN} = 5.5\ {\rm yrs} \left(\frac{v_{\rm SN}}{10^4\ {\rm km/s}}\right) \left(\frac{v_w}{10\ {\rm km/s}}\right)^{-1} \left(\frac{t}{2\ {\rm days}}\right).
\end{equation}
Thus $t_{\rm SN}$ is on the order of the recurrence times for very high mass WDs \citep{Yaron05, Shen09}, so even supernovae with exceptionally early radio observations such as SN 2011fe could hide significant CSM if it were in a SyRNe system. Of course, there are many other lines of evidence that rule out such such a progenitor for SN 2011fe \citep{Bloom12, Chomiuk12, Horesh12, Margutti12, Nugent11, Li11}.

From Figure \ref{fig:ncol}, the location of the nova shell at $t_{\rm rec}$ gives an upper limit on the timescale of interaction with the innermost nova shell of
\begin{equation}
t_{\rm inner} =  12\ {\rm days} \left(\frac{R_s}{10^{15}\ {\rm cm}}\right) \left(\frac{v_{\rm SN}}{10^4\ {\rm km/s}}\right)^{-1}.
\end{equation}
This timescale can vary greatly due to the variability in $R_s$, and we note it is consistent with the $22$ day and $60$ day brightenings observed in the light curve of SN 2002ic \citep{Hamuy03,Wood-Vasey06}.

\section{Conclusions}
The origins of Type Ia supernovae continue to be debated, with some showing evidence of a single-degenerate progenitor \citep{Dilday12, Simon09, Patat07}, and others with strong evidence of a double-degenerate progenitor \citep{Bloom12, Chomiuk12, Horesh12, Margutti12, Nugent11, Li11}. There is thus mounting evidence of multiple channels to a Type Ia.

Decelerating nova shells in SyRNe can have velocities consistent with the few SNe Ia with CSM detections. Nova shells are thin and have high density contrasts ($\sim 100 \times$) compared to the ambient medium, which are important for calculating atomic populations. Additional work on simulating the ionization states of these shells and subsequent radiative transfer during a supernovae is necessary in order to make detailed comparisons to specific spectra. Supernovae in SyRNe systems could be detected both via time-dependent absorption lines in previous shells,  and via rebrightening of the light curve as supernova ejecta hits the shells. 

We note some shortcomings of this model for interacting Type Ia's. Firstly, novae (and RS Oph in particular) are known to have asymmetric ejecta \citep{Hjellming86, OBrien06, Rupen08}, which is not accounted for in our model. An asymmetric outburst (and/or wind profile) can change the ratio of momentum in the wind to that in the ejecta, and thus the coasting velocities. The general picture of a decelerating shell governed by momentum conservation remains, with details such as the density contrast and coasting velocity depending on orientation. 
%It is a careful balance to get a scenario where the WD accretes mass slowly enough to have recurrent nova outbursts, but fast enough to reach $M_{\rm ch}$ before the RG stops ejecting mass. In addition, most of the observations of interacting Type Ia's point to larger CSM masses than can be obtained from a typical RG wind ($\sim 1\ M_\odot$), perhaps indicating a wind from an AGB star. Finally, observations of proposed SNe Ia progenitor systems such as RS Oph have been unable to clearly determine whether the WD is composed of C/O (and could be a Ia candidate), or O/Ne/Mg (which would be an accretion induced collapse candidate).

We thank Ben Dilday, Andy Howell, Sterl Phinney, and Jeno Sokoloski for useful discussions and clarifications. This work was supported by the National Science Foundation under grants PHY 11-25915 and AST 11-09174.

\bibliographystyle{apj}
\bibliography{ms}

\end{document}